\documentclass[aps,pra,floatfix,showpacs,showkeys,epsfig,graphics]{revtex4}%
\usepackage{graphicx}
\usepackage{amsmath}
\usepackage{amsfonts}
\usepackage{amssymb}

\newcommand{\newc}{\newcommand}
\newc{\beq}    {\begin{equation}}
\newc{\eeq}    {\end{equation}}

\newc{\beqa}    {\begin{eqnarray}}
\newc{\eeqa}    {\end{eqnarray}}
\newc{\bs}    {\section}
\newc{\no}    {\\ \nonumber}

\newc{\st}    {\stackrel}

\begin{document}
\title{  Is Dark Matter a BEC or Scalar Field?}
\author{Jae-Weon Lee}\email{scikid@gmail.com}
\affiliation{Department of energy resources engineering, Jungwon university,
             5 Dongbu-ri, Goesan-eup, Chungcheongbuk-do
              367-805, Korea}
\date{\today}
\begin{abstract}
This is a brief review on the history of the Bose-Einstein
condensate (BEC) or boson star model
 of galactic dark matter halos, where ultra-light scalar dark matter particles
condense in a single BEC quantum state. The halos can be described
as a self-gravitating, possibly self-interacting, coherent scalar
field.
 On a scale larger than galaxies,
dark matter behaves like cold dark matter while below that scale
the quantum mechanical nature suppresses the dark matter
 structure formation due to the
  minimum length scale determined by the mass  $m\st{>}{\sim}10^{-24} eV$ and the self-interaction
   of  the particles.
This property could alleviate the cusp problem
and missing satellite problems  of the $\Lambda$CDM model.
Furthermore, this model well  reproduces the observed rotation curves of spiral and dwarf galaxies,
which makes the model promising.
\end{abstract}

\maketitle

Dark matter (DM) and dark energy~\cite{DMreview} are two of the
most important unsolved puzzles in modern physics and cosmology.
Although it is well known that the flatness of the  galactic
rotation curves implies the presence of invisible dark matter
around galactic halos ~\cite{dark},
only a few properties of the dark matter are
known so far. Identification of one DM species by a direct detection experiment,
such as Large Hadron Collider (LHC) or DAMA~\cite{dama},
 is not enough to fully solve the dark matter problem,
because there can be multiple species of DM and
 we have to explain the observed abundance  and distribution of DM in the universe.

While the cold dark matter (CDM) with the cosmological constant ($\Lambda$CDM) model   is popular and remarkably successful in explaining the
large-scale structure of the universe, it seems to
encounter problems on the scale of  galactic or sub-galactic
structure. Numerical simulations with the $\Lambda$CDM model usually predict a cusped central density and
too many sub-halos which are in contradiction with observations~\cite{Salucci:2002nc,navarro-1996-462,deblok-2002,crisis}.
Since typical CDM particles, such as WIMPs, are heavy and slow, they have a tendency to clump forming
a smaller structure,
such as a dark matter star or dark matter planet, while
observations indicate that the smallest DM-dominated structure is a dwarf galaxy.
This implies  a natural minimal length scale of DM.
Furthermore, observed spins of  halos seem to be larger than the
predictions of the $\Lambda$CDM  model.
Although the significance of this discrepancy is still controversial, we cannot safely ignore these problems.
Thus, it is desirable to consider an alternative DM candidate
 both playing the role of CDM for the scales larger than a galaxy
and at the same time, suppressing sub-galactic structures.
Fortunately, we already have one, the Bose-Einstein
condensate (BEC) or boson star dark matter model of  halos~\cite{sin1,myhalo}, which
I briefly review  in this paper. (More generally, one can also call the boson
 star model the scalar field dark matter (SFDM)
model of halos, as dubbed in Ref. 10)

In 1992, to explain the observed properties of galactic rotation curves, such as flatness and ripples,
Sin \cite{sin1} suggested that galactic halos
are   astronomical objects in the BEC  of
ultra-light DM particles such as
pseudo Nambu-Goldstone boson (PNGB). In this model, the halos are like gigantic atoms where
cold boson DM particles are condensated in a single macroscopic wave function
 and the quantum mechanical uncertainty principle
prevents halos from
 self-gravitational collapse.
The halo is described by a wave function $\psi(r)$ of
the non-linear Schr\"{o}dinger equation (Gross-Pitaevskii equation (GPE))
with Newtonian gravity:
\beq
i\hbar\partial_t \psi =-\frac{\hbar^2}{2m}\nabla^2\psi+
GmM_0 \int^{r'}_0 dr'\frac{1}{r'^2}\int^r_0 dr'' 4\pi r''^2
 |\psi|^2\psi(r),
\label{sch}
\eeq
where $m$ is the DM particle mass and $M_0$ is the mass of the halo.
This is the first BEC model of galactic dark matter halos.
According to the model, the condensation of DM particles, whose
  huge Compton wavelength
$\lambda_{comp} = {2\pi \hbar}/{mc} \sim 10~pc$, i.e., $m \simeq 10^{-24} eV$,
is responsible for the halo formation.
The formation of DM structures smaller than the Compton wavelength is suppressed by the uncertainty principle.
Later it was shown ~\cite{corePeebles,PhysRevD.62.103517,0264-9381-17-13-101,PhysRevD.63.063506} that this property could alleviate the aforementioned problems
of the $\Lambda$CDM model.
Despite of their tiny mass, BEC DM particles behave like CDM particles~\cite{Matos:2003pe}
during the cosmological structure formation because
their velocity dispersion is very small.
The Newtonian rotation velocity of  stellar objects at radius $r$  in DM halos
is given by
\beq
 v_{rot}(r)=\sqrt{\frac{GM(r)}{r}},
\label{vr}
\eeq
where $M(r)$ is mass within $r$.

In the same year the author and Koh ~\cite{kps,myhalo} generalized Sin's BEC model by allowing a repulsive self-interaction
among dark matter particles in the context of quantum field theory and general relativity.
This is the first boson star (BS)  model of  dark matter halos (boson halos)  in which
the  dark matter composes a giant boson star described by
 a coherent complex scalar field $\phi$,
which has a typical action
\beq
\label{action}
 S=\int \sqrt{-g} d^4x[\frac{-R}{16\pi G}
-\frac{g^{\mu\nu}} {2} \phi^*_{;\mu}\phi_{;\nu}
 -V(\phi)]
\eeq
with a repulsive potential $V(\phi)$ and a
spherical symmetric metric
\beq
 ds^2=-B(r)dt^2+A(r)dr^2+r^2 d\Omega.
\eeq
In the work, a  quartic potential
$V(\phi)=\frac{m^2}{2}|\phi|^2+\frac{\lambda}{4}|\phi|^4$ was used as an example.
In this model, when $\lambda>0$, the repulsive self-interaction among DM particles,
besides the uncertainty principle, can
provide an additional mechanism against gravitational collapse.
It was also pointed out that one can extend the model to oscillatons, boson-fermion stars, and Q-stars,
which were later investigated by others  independently ~\cite{realfield}.
Since the BS model can be reduced to the BEC model  in the Newtonian limit and since the
observed rotation velocity $v_{rot}\sim 10^{-3} c$, where $c$ is the light velocity,
 the BS model predicts    rotation curves very similar to those of the BEC DM halo model.

 To my knowledge, at least for dark matter halos, this BS model is  also  the first
 model to use the coherent SFDM ~\cite{myhalo}.
Using the boson star theory, it was  pointed out that even a tiny self-coupling of the scalar field
could drastically
 change the scales of dark matter halos and increase the allowable range of the
 mass of dark matter particles.
 From the condition that the halo mass, $M_{halo}$, should be smaller than the critical mass of
 a BS, it was found that~\cite{myhalo}
\beq
\lambda^{\frac{1}{2}}(\frac{M_P}{m})^2\st{>}{\sim} 10^{50},
\label{mlambda}
\eeq
which leads to $10^{-24}~eV\st{<}{\sim} m \st{<}{\sim}10^3~eV$ for $0\le \lambda \le1$.
Here, the lower bound came from the Compton wavelength for $\lambda=0$.
This is a relation between the mass and  the coupling of
 the halo dark matter particle. Henceforth, I will designate the two models (BEC and BS models)
 as the BEC/SFDM model for simplicity.

Unfortunately, these two original works  were not widely known.
Later,  similar ideas were  rediscovered and developed  by many authors
using various potentials and fields such as the
 massless field ~\cite {Schunck:1998nq}, the non-minimal coupling ~\cite{PhysRevLett.84.3037},
quintessence~\cite{PhysRevD.64.123528,PhysRevD.65.083514},
the short-range force (Repulsive DM) ~\cite{repulsive},
fuzzy DM~\cite{Fuzzy}, fluid DM ~\cite{corePeebles}, the nontopological soliton  ~\cite{Nontopological},
the $cosh$ potential~\cite{PhysRevD.62.103517,Alcubierre:2001ea}
and many others ~\cite{Fuchs:2004xe,Matos:2001ps,0264-9381-18-17-101,PhysRevD.63.125016,Julien}.
(See Ref. 31  for a brief review.)
However, it is clear that the common concept behind all these DM models, i.e., BEC DM halos
or SFDM halos, was  introduced in
 the BEC model and the BS model as early as 1992.
(Note that a BS had been called by many different names, such as
a soliton star, a scalar star, and a Klein Gordon geon~\cite{0264-9381-20-20-201}.)
On the other hand, although
strongly self-interacting dark matter (SIDM)~\cite{sidm} and the BS model
share a property  that DM particles may strongly self-interact repulsively,
 the two models are  different  in that
DM particles in the SIDM model are not in a coherent state and are usually
very massive.
There is an  idea that a vector field in a modified gravity action could
be identified with a BEC ~\cite{moffat-2006}.

The BEC/SFDM model has many good features. In the original works,  it was emphasized that
this model could explain the flatness and the  ripple structure of the rotation curves,
which was later verified by observations
by many independent researchers
~\cite{PhysRevD.64.123528,0264-9381-17-1-102,Mbelek:2004ff,PhysRevD.69.127502}.
Furthermore, it was later found  that these models could be  free from the cusp problem  ~\cite{corePeebles,Riotto:2000kh}
 and the missing satellite problem  ~\cite{corePeebles,PhysRevD.62.103517,0264-9381-17-13-101,PhysRevD.63.063506},
which made this model  promising.
Recent studies on the dark matter in  halos of satellite dwarf spheroidal (dSph)
galaxies of the Milky Way
have attracted much interest~\cite{gilmore-2008,gilmore-2006}.
The dSph
 galaxies seems to be the smallest dark-matter-dominated astronomical objects and, hence,
 are
 ideal objects for studying the nature of dark matter.
The observation suggest that a
typical dSph never has a
size  less than $R_{halo}\sim 10^3pc$, and
that the central dark matter halo density profile is not cusped.
The boson star model of dSphs seems to be also consistent with
the analysis of Ref.  40.
According to the observations, the halos of dSphs
seem to be cored halos with isotropic stellar velocity, and
more massive halos have higher central densities.
This mass distribution shows a universal profile
and implies  a large spatial scale length.
Interestingly, all these are characteristic of BS halos.
The mass of a boson star is proportional to the central density~\cite{0264-9381-20-20-201} and
a BS has a constant core density~\cite{review}.
 Recently, it was also shown that the BEC (automatically, the BS model, too)  reproduces
 the rotation curves of
  dwarf galaxies very well  ~\cite{PhysRevD.68.023511,Boehmer:2007um}.
  Note that no other DM models has successfully   reproduce the observed
  rotation curves at this level so far.
There are the theories considering the role of visible matter to explain
 the discrepancy between the numerical simulations in the $\Lambda CDM$ model
and the observations,
 but, these explanations seem to be rather ad hoc and complicated
 compared to the explanation of the BEC/SFDM model.

Let us discuss the BEC/SFDM model in detail.
The cold gravitational equilibrium configurations
of a scalar field, called a boson star (See, for example, Refs. 43 and 46), were found by solving
the Klein-Gordon equations with gravity
decades ago~\cite{orig}.
 These configurations
are adequate for a relativistic
 extension of the BEC model.
From the action in Eq. (\ref{action}), dimensionless time-independent Einstein and
scalar wave equations appear as in Ref.  47:
the (tt) and the (rr) components of Einstein equation are
\beq
\frac{A'}{A^2x}+\frac{1}{x^2}[1-\frac{1}{A}]
-[\frac
{\Omega^2}{B}+1]\sigma^2
-\frac{\Lambda}{2}\sigma^4 -\frac{\sigma'^2}{A}=0\\
\eeq
and
\beq
\frac{B'}{ABx}-\frac{1}{x^2}[1-\frac{1}{A}]
-[\frac
{\Omega^2}{B}-1]\sigma^2+
\frac{\Lambda}{2}\sigma^4 -\frac{\sigma'^2}{A}=0,\\
\eeq
respectively, while
the equation of motion for the field is
\beq
\sigma''+[\frac{2}{x}+\frac{B'}{2B}-\frac{A'}{2A}]\sigma'
+A[(\frac{\Omega^2}{B}-1)\sigma -\Lambda \sigma^3]=0,
\label{eqs}
\eeq
where $x=mr, \Omega=\frac{\omega}{m}$ ,
$A\equiv[1-2\frac{M(x)}{x}]^{-1}$, $
\phi(x,t)=(4\pi G)^{-\frac{1}{2}} \sigma(r)
e^{-i\omega t}$, and $\Lambda=
\frac{\lambda m^2_p}{4\pi m^2}$.
In Refs. 8 and 9    the Newtonian $v_{rot}$ was used while
the  relativistic rotation velocity includes the contribution from the pressure~\cite{0264-9381-20-20-201}:
\beq
v_{rot}=\sqrt{\frac{x B'(x)}{2B(x)}}.
\eeq

Figure 1 shows rotation velocity curves
for the case with $\Lambda=300$.
The parameters are $B(0)=0.780$ and $\sigma(0)=0.01$.
Only the DM contribution is considered here.
Including the visible matter  changes the slope
of the curves  and  explains well the variety of
 observed galaxy rotation curves~\cite{sin2}.
Note that the density $\sim \sigma^2$ shows the core shape distribution.

\begin{figure}[htbp]
\includegraphics[width=0.4\textwidth]{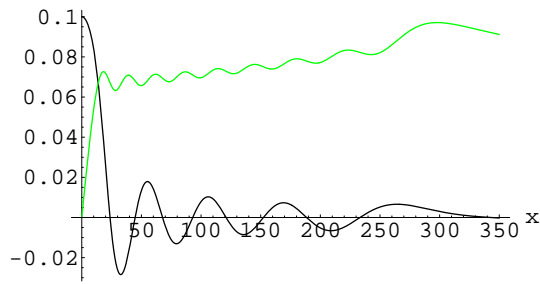}
\caption{
Rotation velocity (green) and $10\times \sigma$
 as a function of position $x$
for an 8 node solution.
The parameters are $\Lambda=300$, $\Omega=0.9$,
 $B(0)= 0.780 $ and $ \sigma(0)=0.01$.}
 \label{flat}
\end{figure}

For DM particles to be in the BEC state, they have to be  gauge singlets,
and the BEC phase transition must happen at the early universe.
Regarding the origin of the DM scalar particles, it is interesting~\cite{PhysRevD.64.123528}
  that the observed rotation velocity, $v_{rot}\sim 10^{-3}$ (in units of $c$), corresponds
to the central field strength of about GUT or inflation scale
$\sigma(0)\sim v_{rot}^2 M_P$.
This field value may be related to the BEC phase transition~\cite{Yu:2002:0264-9381:L157}.
Figure \ref{flat}  shows $\sigma$ and rotation velocity curve of
an 8 node solution($n=9$). The zero node solutions seem to be adequate for dwarf galaxies
 while higher-node solutions are required for  larger galaxies.

For free-field case  ($\Lambda=0$) \cite{free}, it was found that
there is a maximum  mass
 $ M_{max}=0.636\frac{M^2_p}{m}$
for the zero-node solution.
For the case  $\Lambda\neq 0$, a new scale appears because of the
  repulsive force preventing the halo from gravitational
 collapse.
In this case, the typical length scale
 is $R\sim \Lambda^{\frac{1}{2}} /m$,
thus, the typical mass scale is $\frac{R}{G}
\sim \Lambda^{\frac{1}{2}} M_P^2/m$.
A numerical study\cite{colpi} shows
 $M_{max}=0.236 \Lambda^{\frac{1}{2}}\frac{M^2_p}{m}$
for zero-node solutions.
Note that
 $\Lambda= \lambda M_P^2/4\pi m^2$
is very large even for very small $\lambda$
 due to the smallness of $m$
 relative to $M_P$; hence, the
self-interaction effect is non-negligible.
The Newtonian limits of the last equations for $\Lambda=0$ lead to
the non-linear Schr\"oedinger equation for
  Sin's model~\cite{newt1}.

A remaining concern about the BEC/SFDM model may be the stability of the excited states
needed to explain flat rotation curves for large galaxies.
It was shown ~\cite{sin1,myhalo} that the
DM halo is stable enough against  gravitational
radiation~\cite{gw}, and the evaporation and collapse procedure~\cite{sin1}.
However, there are arguments~\cite{gcooling2} that the exited states are not stable against  gravitational cooling
~\cite{gcooling}.
This difficulty may be alleviated by considering the rotation of the halos.
Finite angular momentum may prevent the collapse of the excited states, as is usual in
an astronomical object.
According to the bottom-up structure formation scenario of
the $\Lambda$CDM model, large galaxies were made by the
merging of smaller galaxies.
Thus, it is plausible that the halos of the large galaxies got angular momentum during the merging process
 and have vortices within them~\cite{PhysRevD.55.6081}.

Another merit of the BEC/SFDM model is that collisions of halos ~\cite{collision1} can be described as
BS collisions ~\cite{choi1},
which could open an intriguing new field of galaxy dynamics and evolution study.
To do this, we need the Newtonian approximation of the Einstein equation
and the
scalar field equation
 \beq
 \left\{ \begin{array}{l}
 \nabla _{} ^2 V = \sigma ^2  + \frac{{\Lambda \sigma ^4 }}{4} \\
  \nabla ^2 \sigma  = 2V\sigma    \\
\end{array} \right.    \\
 \eeq
where $V$ is the Newtonian gravity potential.

The BEC/SFDM model also provides us a fascinating possibility
of simulating  evolution of DM objects by using the results obtained from
BEC experiments in a laboratory.
It will also be interesting to study
the cosmological effects
 of  more general scalar fields \cite{rubin-2001-92}.

In conclusion,
since the BEC/SFDM model has passed test by test
and seems to overcome the difficulties of heavy CDM particles,
the future of this model seems to  be very bright.
Especially, if  no WIMP is found at LHC or in
other ground-based experiments soon, this model could be an interesting  alternative.
\\
\\

\section*{ }
This work was supported in part by the topical research program (2008 -T-X)
of Asia Pacific Center for Theoretical Physics.


\vskip 5.4mm
\newpage


\end{document}